\documentclass[onecolumn,showpacs,preprintnumbers,amsmath,amssymb]{revtex4}
\usepackage{graphicx}
\usepackage{dcolumn}
\usepackage{bm}

\newcommand{\be}{\begin{equation}}
\newcommand{\bea}{\begin{eqnarray}}
\newcommand{\ee}{\end{equation}}
\newcommand{\eea}{\end{eqnarray}}
\begin{document}

\title{Computational Models For Epilepsy}

\author{Roxana A. Stephanescu}
\affiliation{Department of Pediatrics, Division of Neurology} \altaffiliation{University of Florida, Gainesville, 32611, USA}

\author{Shivakeshavan R.G.}
\affiliation{Department of Pediatrics, Division of Neurology\\
and\\ Department of Biomedical Engineering}
\altaffiliation{University of Florida, Gainesville, 32611, USA}

\author{Sachin S. Talathi}
\affiliation{Department of Pediatrics, Division of Neurology\\
and\\ Department of Biomedical Engineering and\\ Department of Neuroscience}
\altaffiliation{University of Florida, Gainesville, 32611, USA}
\altaffiliation{Corresponding author}
\email{talathi@ufl.edu}

\date{\today}
\clearpage

\begin{abstract}
Epilepsy is a neurological disease characterized by recurrent and spontaneous seizures. It affects approximately 50 million people worldwide. In majority of the cases accurate diagnosis of the disease can be made without using any technologically advanced techniques and seizures are controlled using standard treatment in the form of regular use of anti-epileptic drugs. However, approximately 30\% of the patients suffer from medically refractory epilepsy, wherein seizures are not controlled by the use of anti-epileptic drugs. Understanding the mechanisms underlying these forms of drug resistant epileptic seizures and the development of alternative effective treatment strategies is a fundamental challenge in modern epilepsy research. In this context, the need for integrative approaches combining various modalities of treatment strategies is high. Computational modeling has gained prominence in recent years as an important tool for tackling the complexity of the epileptic phenomenon. In this review article we present a survey of different computational models for epilepsy and discuss how computer models can aid in our understanding of brain mechanisms in epilepsy and the development of new epilepsy treatment protocols.
\end{abstract}
\maketitle

\section{Introduction}

Epilepsy is a neurological disease that pervades through a wide spectrum of human lifestyles. The disease is primarily characterized by patients suffering from recurrent bouts of spontaneously occurring seizures. These seizures not only disrupt normal living but can also cause mental and physical damage and in extreme cases, even death. Attempts to treat epilepsy usually comprise of medication, brain stimulation, surgery, dietary therapy or various combinations of the above, directed toward the goal of eliminating or controlling seizures \citep{Shorvon}. For many epileptic patients, seizures are well controlled with anti-epileptic drugs (AEDs). However, approximately 30\% of the patients suffers from medically refractory epilepsy, wherein seizures continue to occur despite treatment with a maximally tolerated dose of a first-in-line AED or in combination with at least one adjuvant medication \citep{Remy2006}. This has motivated clinicians and researchers alike to investigate and understand the sources of seizures in refractory epilepsy using techniques from varying scientific disciplines such as molecular biology, genetics, neurophysiology, neuroanatomy, brain imaging and computer modeling.

Certainly, computational modeling of epileptic brain activity is not the first approach that one might consider to solve the problem of refractory epilepsy. Experimental research has identified many factors involved in the generation of epileptic seizures \citep{Schwartzkroin1993,Holmes1997}. The findings suggest that the etiologies of refractory epilepsy are so diverse and complex, that it is a formidable task to obtain a single framework  that categorizes all the pathophysiological changes in genetic, molecular, cellular and neuronal network level properties of the brain involved in uncontrolled recurrent epileptic seizures. As a result, at the first glance, it is difficult to understand how computational models can aid in unravelling the complexity of the epileptic syndrome, translate into rehabilitation or aid in the development of new treatment strategies.

Computational modeling have a long and rich history of enabling better understanding of normal brain function \citep{Traub1994,Markram2006}.  Computational models have also found applications in  understanding the pathologic behavior of brain networks \cite {Santhakumar2005,Traub2005,Kudela1997,Kudela2003,Kudela2003a}. A strong case therefore be made for developing computational models for epilepsy not as a replacement for other approaches, but as a mean to obtain new insight into pathogenesis and treatment of epilepsy\citep{Lytton2008}. Models for epilepsy have found application in seizure prediction \citep{OSullivan-Greene2009}, probing the cellular and network mechanisms of seizures \citep{Bazhenov2008} and also as a tool to guide strategies for therapy by surgical, pharmacological and electrical stimulation techniques \cite{Iasemidis2003}. 

Computational models provide a unique framework to integrate data from experimental findings to foster new hypotheses, which in turn can guide future experiments. Models of epilepsy provide excellent tools and techniques to relate variables across multiple levels of analysis, thereby offering the possibility of establishing links between the hierarchy of brain networks involved in the origin and spread of epileptic seizures. Another significant advantage of modeling is that experiments that are more challenging to perform can be easily simulated through a computer model. This is particularly true in the study of neurodegenerative diseases such as epilepsy since it is relatively easy to mimic lesions caused by brain injury to determine the underlying mechanism of disease progression caused by the lesion. There are few practical barriers and virtually no ethical barriers to conducting a large number of exploratory virtual experiments. This can allow researchers to perform a large number of virtual experiments in order to extract the most relevant information to be further verified in a lab-experimental setting. As a result the last decade has witnessed the emergence of a wide variety of computational models of epilepsy \citep{Lytton2008}.

In this paper we present a brief survey of computational modeling approaches in modern epilepsy research.  We first introduce basic concepts from the mathematical theory of dynamical systems in support of the idea that epilepsy can be understood as a dynamical disease \cite{LopesdaSilva2003,Milton2010}. We then review the most commonly adopted computational frameworks for modeling epileptic brain networks with special emphasis on case examples from recent literature that highlight the various modeling approaches adopted by epilepsy researchers to tackle the complexity of the epileptic syndrome. This discussion is followed by a brief review on  the utility of computational models in developing novel epilepsy treatment protocols. Here, we present some preliminary results from our group using light stimulation based feed-back control strategies to regulate pathological brain activity. We conclude with a discussion on the future prospects for computational models  in developing a better understanding and novel therapy protocols for epilepsy.

\section{Understanding the dynamical characteristics of epilepsy}
As mentioned earlier, epilepsy is considered to be a dynamical disease \cite{LopesdaSilva2003,Milton2010}. The transition of the brain to an epileptic state is a time-evolving process that begins with an incidence of a precipitating brain injury such as stroke, trauma or status epilepticus. Temporal events ranging from millisecond time scale (neuronal action potentials) to hours and several days (hebbian plasticity) participate in the eventual transition of the brain to spontaneously seizing state. Seizures themselves are temporal events lasting several tens of seconds and ending abruptly. This kind of temporal evolution pattern is also observed in other systems such as earthquakes and weather patterns and has been the object of mathematical and computational analysis for the last several decades.

From dynamical systems point of view, the brain can be considered as a multi-dimensional dynamical system defined via an independent set of {\it system variables} such as the neuronal membrane potentials, which evolve in time following a set of deterministic equations and {\it system parameters} that either do not evolve in time (for example, conductance of ion channels on neuronal membrane) or their evolution happens on much slower time scales relative to the evolution of system variables (the strength of synaptic conductances, which follow hebbian plasticity rules).



In order to illustrate the key concepts from dynamical systems that are essential to understand the notion of epilepsy as a dynamical disease, we consider an example of a 1-dimensional dynamical system, defined via a system variable $x$. The time evolution of the dynamical system is given as:
\begin{equation}
\dot{x} = f(x) + I;
\end{equation}
where $\dot{x}=\frac{dx}{dt}$ measures the rate of change in $x$ as a function of time $t$. The function $F(x)=f(x)+I$, referred to as the {\it vector field}, governs the time evolution $\phi^{t}(x_{0})$ of $x$  for a given value for system parameter $I$ and initial condition $x(t=0)=x_{0}$. $\phi^{t}(x_{0})$ is referred to as the {\it flow} or {\it trajectory} of the dynamical system.  {\it Fixed point} equilibrium is defined as the value of $x=x_{s}$, at which that the vector field $F(x_{s})=0$. In this case, there is no flow in the system, i.e. $\phi^{t}(x_{s})=x_{s}$. The stability of the fixed point equilibrium state depends on the value for the first derivative $\lambda=\frac{\partial F}{\partial x}$ evaluated at the fixed point $x_{s}$. If $\lambda <0$, the fixed point is considered as a stable (or an attractor) equilibrium point, where as if $\lambda >0$, the fixed point is unstable (or repeller). If $\lambda=0$, a second derivative $\frac{\partial^{2}F}{\partial x^{2}}$ must be employed to determine the stability nature of the equilibrium state in question. It is important to note that both the fixed points as well as their stability depends on the values of the system parameter $I$. As we further explain below, by changing the system parameter value, the stability of the fixed point can be influenced. A bifurcation occurs when changes in value of the system parameters change the stability properties of the equilibrium point. These concepts and definitions can be easily extended for a multi-dimensional dynamical system with the system variables $x\in \mathbb{R}^{n}$, where $n$ is the dimension of the dynamical system.  The time evolution of each system variable is governed by its own vector-field: $\dot{x_{i}}=F_{i}$, $i=1\cdots n$. The flow of the system is now described in an n-dimensional phase-space (a multi-dimensional space with axes defined by the system variables), $\phi^{t}(x_{1}(0),\cdots, x_{n}(0))$ and the stability of the equilibrium states is obtained by identifying the eigenvalues of the Jacobian matrix $J$ with matrix elements $J_{ik}=\frac{\partial F_{i}}{\partial x_{k}}$ \cite{Strogatz}.

We further illustrate the key concepts presented above by presenting a specific example of a dynamical systems based two-dimensional neuron model \cite{Morris:1981dz}. In Figure 1, we show the phase space for the neuron model. The system variables for the model are the neuronal membrane potential $v$ and the gate variable for the potassium ion channel $w$. The model equations are as described below:
\begin{eqnarray}
\frac{dv}{dt} &=& F_{1}(v,w) \nonumber \\
\frac{dw}{dt} &=& F_{2}(v,w)
\end{eqnarray}
where \begin{eqnarray}
F_{1}(v,w)=\frac{1}{C}\left(-0.5g_{Ca}\left[1+\tanh\left(\frac{v-V_{1}}{V_{2}}\right)\right](v-V_{Ca}) - g_{K}w(v-V_{K}) - g_{L}(v-V_{L}) + I\right)\nonumber
\end{eqnarray} $$F_{2}(v,w)=\phi\cosh\left(\frac{v-V_{3}}{2V_{4}}\right)\left(0.5\left[1+\tanh\left(\frac{v-V_{3}}{V_{4}}\right)\right] - w\right)$$ The system parameters $C=20$ $\mu$F/cm$^{2}$,$g_{Ca}=4$mS/cm$^{2}$, $V_{Ca}=120$ mV, $g_{K}=8$ mS/cm$^{2}$, $V_{K}=-80$ mV,$g_{L}=2$ mS/cm$^{2}$, $V_{L}=-60$ mV, $V_{1}=-1.2$ mV, $V_{2}=18$ mV, $V_{3}=2$ mV, $V_{4}=8.3$ mV remain unchanged under a typical experimental paradigm. The system parameter $I$ is the external current that can be injected into the neuron model and is under the control of the experimenter. The neuronal dynamics represented schematically in the phase space representation in Figure 1a is for a specific instance when $I=35$$\mu$A/cm$^{2}$. As explained in the previous paragraph, the fixed point equilibrium state $\{v_{s},w_{s}\}$ for the above neuron model satisfies the constraint $F_{1}(v_{s},w_{s})=0$ and $F_{2}(v_{s},w_{s})=0$. The functions $F_{1}(v,w)=0$ (pink curve in Figures 1a and 1b) and $F_{2}(v,w)=0$ (orange curve in Figures 1a and 1b), are referred to as the {\it nullclines} and intersection points of the two nullclines correspond to the fixed point equilibrium states of the system. We find that there are three fixed point equilibrium states (labeled as black, cyan and red dots in Figure 1a). Stability analysis shows that there is a one stable fixed point (black dot in Figure 1a) and two unstable fixed points (cyan and red dot in Figure 1a). The set $j=1\cdots 8$ of blue curves in Figure 1a are the trajectories $\phi^{t}(v^{j}_{0},w^{j}_{0})$ starting from different initial conditions $\{v^{j}_{0},w^{j}_{0}\}$ that evolve towards the fixed point attractor equilibrium state in the phase space. The set of all initial conditions from which trajectories evolve to a given fixed point attractor define the {\it basin of attraction} for that attractor. Biophysically the stable fixed point equilibrium represents the resting state of the neurons membrane potential. We also see that there are trajectories that move away from the fixed point repeller (red dot in Figure 1a) and merge into a closed-loop trajectory (shown in black in Figure 1a), which is referred to as the {\it limit cycle} attractor. Limit cycles correspond to the case when neurons are in a periodic spiking state. For most dynamical systems the fixed point is the default attractor for the system. However, many of the stable states in a higher dimensional dynamical system may exhibit attractors that are not fixed points. Another important feature of dynamical systems, evident from Figure 1a is multi-stability. Multiple attractors can reside in the phase space at the same time and depending on the initial conditions, the dynamical system can evolve to any one of these attractors. For the example considered, the dynamical system has two attractors: one stable fixed point and one limit cycle. A simple computational principle that follows from the existence of the above two stable attractors is that, the neuron can switch from resting state to periodically spiking state via an appropriately timed single pulse of current that moves the dynamics of the neuron in the phase space from a point closer to the fixed point attractor to a point closer to the limit cycle attractor and vise-versa. Finally, a very important concept in dynamical systems is the concept of {\it bifurcation}. It designates a qualitative change in the dynamical behavior of the system associated with modifications in the system parameters. The precise values of the parameters for which this change occur are called {\it bifurcation points}. In Figure 1b, we show the phase space for neuron model when $I=30$$\mu$A/cm$^{2}$. We see a qualitative difference in the system dynamics as compared to the case considered in Figure 1a. This observation suggests that for particular value of system parameter $I=I^{*}$, where $30<I^{*}<35$ there is a bifurcation in the system dynamics. For the case $I=30$$\mu$A/cm$^{2}$, the system still exhibits three fixed point equilibrium states however  there is only one attractor. All the trajectories evolve towards the fixed point attractor. The neuron in this case is not able to generate periodic spiking behavior.

\begin{figure*}[htbp]
  \centering
    \includegraphics[scale=0.5]{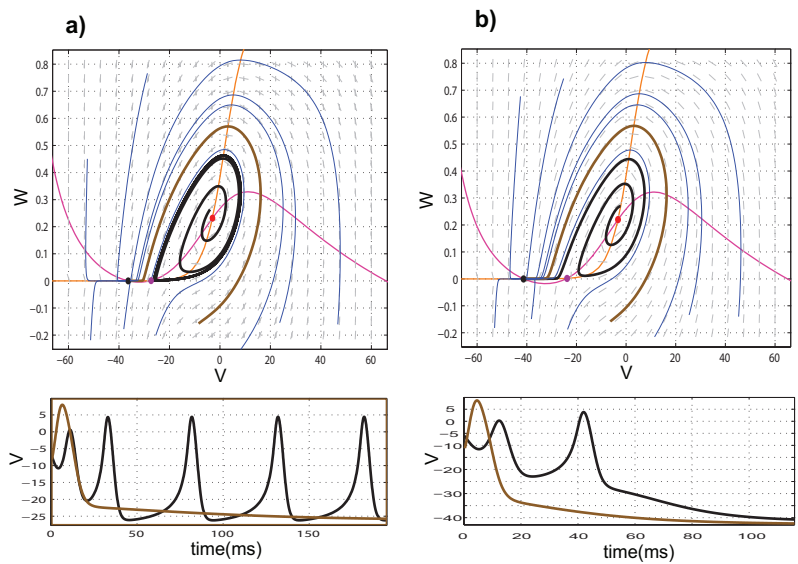}
  \caption{ a) Phase space representation of neuronal dynamics is shown when $I = 35$ $\mu$ A/cm$^{2}$. Several dynamical quantities are exemplified: the nullclines, the stable (black dot) and unstable fixed equilibrium points (brown and red dots) and trajectories representing the time evolution of neuron model system variables . Two examples trajectories are emphasized: a trajectory evolving towards the limit cycle (black curve) and the second trajectory evolving towards the fixed point attractor (brown curve). The voltage time series associated with these trajectories are displayed in the panel below.  b) Phase space representation of neuronal dynamics is shown when $I = 30$ $\mu$ A/cm$^{2}$. Some features presented in a) change via a bifurcation generated by the changes in the system parameter $I$. The limit cycle is no longer an attractor in the phase space; All trajectories evolve towards the only stable fixed point attractor in the network.}
  \label{Fig1}
\end{figure*}

With these concepts in mind we can now understand how epilepsy can be viewed as a dynamical disease \cite{LopesdaSilva2003,Milton2010} in the sense of transitions from a ``normal" brain state attractor, to a ``pathological" brain state attractor in an epileptic brain network. Possible mechanisms include sudden transitions from one attractor to another in a multi-stable dynamical landscape by means of changes in certain critical system parameters (bifurcation) or perturbations induced by noise in the system. These transitions could be mediated by various modulatory mechanisms active in the brain or by triggers originating outside of the central nervous system (for instance light induced seizures). Based on this view, models for epileptic brain networks susceptible to epileptic seizures have been proposed. For instance, in \cite{LopesdaSilva2003} three models are proposed to account for a broad spectrum of epilepsies. The first model suggests that the attractors for the ``normal"  and the ``pathological" brain states are very close in the phase space of the brain dynamics prone to epileptic seizures as compared to brain dynamics that are not susceptible to epileptic seizures; therefore, random fluctuations of some system parameters are sufficient to induce a transition to the pathological seizing state. In this scenario, seizure occurrence may not be predictable. In the later two models, the authors propose that these attractors are further away such that random fluctuations in system parameters cannot trigger a seizure; instead, the system is characterized by unstable system parameters that are very sensitive to endogenous and/or exogenous factors. These parameters may gradually evolve in time in
such a way that the basins of attraction corresponding to the ``normal" and ``pathological" brain attractors  get closer in the phase space and any random fluctuation can then facilitate the transition to seizures. It is plausible that the gradual evolution of system parameters are detectable in the EEG by means of signal analysis techniques and algorithms. As a result, this mechanism for the brain transition to epilepsy offers the means to anticipate and design appropriate treatments protocols to evade an impending seizure.

\section{Modeling attempts in epilepsy}
Very often, when attempting to construct a model for a complex brain disease such as epilepsy, one is confronted with an overwhelming amount biological details, which if one were to attempt to model in its entirety, would rapidly exceed the theoretical and computational resources that may be available to the modeler. A key question for any modeler is what trade off can he/she make in simplifying the process of model development, while maintaining certain degree of biological realism relevant to  the questions of interest? Based on the level of simplification, several classes of models become available. In the following, we will discuss models for epilepsy that follow two of the most commonly employed criteria of simplification: (a) the type of the model i.e., deterministic vs. non-deterministic and (b) the spatial scale of modeling i.e., micro vs. macro.

\subsection{Deterministic models for epilepsy}
Deterministic models for epilepsy are usually presented in the form of system of ordinary differential equations (ODEs) (of the form given in eq. 1 and eq. 2). These models assume that the time evolution of system variables are completely governed by the set of ODEs. In other words, if initial conditions and the system parameters are specified, one can evaluate the state of the system at any time in the future. Due to the high degree of structural and temporal complexities of an epileptic brain, many deterministic models aim to represent the dynamics of an epileptic brain by limiting the analysis to a given spatial scale of resolution such as the micro-scale, which is typically confined to neuronal networks within a given brain region such as the hippocampus vs. macro-scale, wherein attempt is made to model the dynamics of ensemble of neuronal populations involving multiple brain regions.

On the micro-scale, modelers are concerned with questions related to the dynamical behavior of individual neurons including pathology in neuronal ion channels, contribution of neuronal morphology (dendritic tree, axonal arborization) and interaction between neurons and its surrounding environments. Many of the epilepsy models that fall under this class, adopt the Hodgkin-Huxley framework of conductance based-neuronal modeling \cite{hodgkin1952}, with equations of the form described for neuron model in eq. 2. The spatial structure of the neuron is ignored, rather the neuron is considered as an uni-compartment system representing the soma. At the single neuron level, attempts to explain the mechanisms of enhanced excitability in an epileptic brain focus on the changes in the kinetic properties of individual ion channels (channelopathy) that comprise the neuronal membrane. For example, several computational models for a single CA1 pyramidal neuron have examined the role for up regulation of the I$_{\text{h}}$ currents as a potential source of pro-excitability in epileptic hippocampal  networks \cite{Dyhrfjeld-Johnsen:2008uq,Golding:2001fk}. 

Yet other studies have focused on the role for neuronal morphology and their contributions to increased excitability observed in epileptic brain networks. Compartmental modeling represents the most general framework for neuronal modeling efforts that attempt to address the contribution of the spatial extent of the neuron to an epileptic brain \cite{Rall:1967qa,Mainen:1996ij}. It represents the highest level of detail for constructing neuron models. The idea is to divide the neuron into compartments such that electrical potential within each compartment is assumed to be constant. Multi-compartment models of hippocampal pyramidal neurons have been constructed with the number of compartments ranging from a couple \cite{Pinsky1994,Kamondi1998} to several hundreds \cite{mcompart1,mcompart2}. As explained later in this section, such multi-compartmental models have been employed in several applications to investigate the contributions from the spatial extent of pyramidal cells to enhanced excitability in epileptic brain networks.

The above approach of conductance based compartmental modeling can become computationally expensive, especially to simulate networks comprising thousands of neuronal units. In order to tackle the issue of computational complexity, various simplified neuronal models are available (an example was presented in eq.2). These models are specifically designed to reduce the complexity of the system to be modeled, i.e., reduce the number of ODEs required to model a neuron, while still reproducing important dynamical characteristics observed under varying experimental conditions.  Moris-Lecar \cite{Tass:2007vn,Volman:2011ys}, Hindmarsh-Rose \cite{Percha:2005zr} and FitzHugh-Nagumo  \cite{Balazsi:2001kx}  are some of the most prominent neuron models under this category that have been employed to investigate dynamics in epileptic brain networks.

\begin{figure}[htbp]
  \centering
    \includegraphics[scale=0.7]{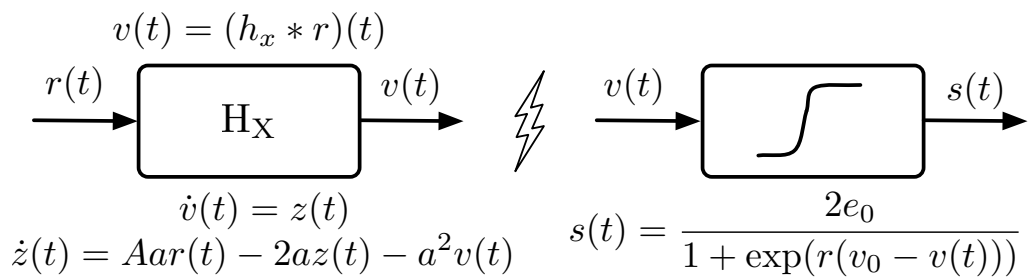}
  \caption{: Schematic block diagram representation of the mean field neural population model. The linear transfer function block converts average presynaptic firing rate of neural population $r(t)$ to average postsynaptic membrane potential (PSP) $v(t)$. The nonlinear sigmoidal function block converts the mean PSP $v(t)$ into the mean postsynaptic spiking rate $s(t)$ of neural population. Mathematical operations performed by each block are also shown.
}
  \label{Fig2}
\end{figure}

On the macro-scale, the majority of the deterministic models attempt to model the dynamics of populations of neurons across brain  regions rather than modeling membrane potential dynamics for individual neuron in a given brain region.  Given that many of the experimental techniques for the measurement of epileptic brain activity such as electroencephalograms and field potential recordings represent large populations of neurons across multiple brain regions,  models at this scale are suitable for direct comparison with experimental data. Wilson and Cowan pioneered the macro-scale modeling approach in a series of theoretical papers in the seventies \cite{WilsonCowan1,WilsonCowan2}. The basic idea is to model the averag firing activity of the ensemble of neuronal populations. In general, the temporal activity of a subpopulation of neurons in the network is modeled via an impulse response function of the form $H(t)=Aate^{-at}$, which models the mean synaptic activity $V(t)=(H\ast r)(t)$ generated by the subpopulation with mean firing rate $r(t)$ and a nonlinear sigmoidal function of the form $S(t)=\frac{2e_{0}}{1+e^{r_{0}(v_{0}-V(t))}}$, which transforms the synaptic activity in the network into mean firing rate for neurons within the subpopulation \cite{Lopes-da-Silva:1974kl}. The system parameters are $A$, which represents the maximum subpopulation synaptic activity, $a$, which corresponds to the lumped representation of the sum of reciprocal time constant of the passive cell membrane and all the spatially distributed delays in the network, $e_{0}$ which represents the maximum firing rate within the subpopulation, $v_{0}$ is the mean firing threshold and $r_{0}$ mimics the gain in synaptic activity. The schematic for modeling firing activity of neuronal subpopulation described above is shown in Fig. \ref{Fig2}.

\begin{figure}[htbp]
  \centering
    \includegraphics[scale=0.5]{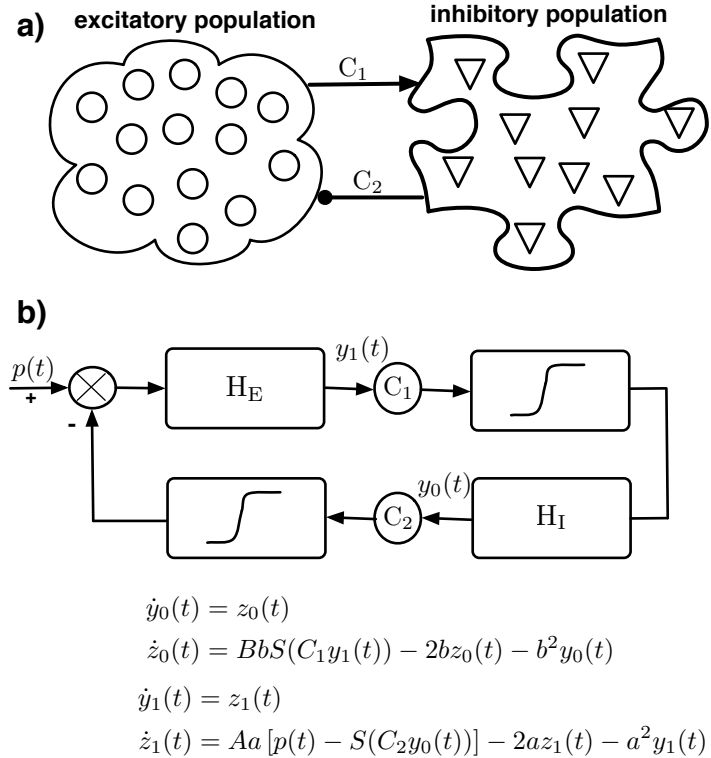}
  \caption{:  a) Schematic diagram of two interacting neural populations. b) The neural mass representation of the interaction: $y_{0}(t)$ and $y_{1}(t)$ represent the average postsynaptic membrane potentials (PSPs) of the inhibitory and excitatory neuron populations respectively. The parameters (A,a) and (B,b) model the maximal PSP amplitudes and the time constants of the excitatory and inhibitory transfer function blocks respectively. $p(t)$ represents the pulse density of neighboring or distant neural populations that synapse onto the excitatory neural population block, and can be modeled as an arbitrary function including Gaussian white noise. The interaction between the two neural population blocks can be modeled using set of 4 nonlinear ODEs.}
  \label{Fig3}
\end{figure}

The simplest network models constructed within this framework include interaction between single subpopulations of excitatory and inhibitory neurons as shown in Figure \ref{Fig3}. The constants $C_{1}$ and $C_{2}$ account for the total number of synapses between the two neural populations. An extension of this approached has been implemented in Jansen-Rit model \cite{JansenRit} to explain alpha rhythm generation within cortical columns and recently in Wendling et al. \cite{Wendling2002} to explain epileptic fast activity by means of impaired GABAergic dendritic inhibition.

\subsubsection{Case Examples}
We will discuss three case examples of deterministic computational models, which highlight the utility of various modeling frameworks discussed above in exploring different aspects of brain network mechanisms in epilepsy. The choice of these case examples is motivated by the fact that (a) they exemplify  commonly adopted micro- and macro-modeling frameworks in computational modeling of epilepsy and (b) they demonstrate how the choice of modeling hierarchy is governed by the question that the study attempts to address.  
The first case example is the work of Ullah et al \cite{Ullah2009}. The authors investigate the conditions under which  persistent neural activity, defined as spatially restricted sustained firing activity in neuronal networks in response to a brief input stimulus, can transition to seizure-like activity, defined as the condition wherein the spatially restricted neural activity spreads to the entire network. The authors focus on the role for extracellular space and glia in modulating the excitatory-inhibitory balance in the network, which in turn influence the stability and the spread of persistent network activity. The modeling framework in this study is based on a generic micro-scale neuronal network model of excitatory and inhibitory Hodgkin-Huxley neurons that are modified to explicitly model the concentration gradients of intra- and extra-cellular potassium ions as a function of the glial activity and active ionic pumps.

In conditions of excitatory-inhibitory balance and normal glia function, the network is able to maintain spatially restricted persistent neural activity in response to weak stimulation. Further exploration of the network dynamics by the authors suggests a fine balance between the overall levels of excitation and inhibition (defined through values for the excitatory-excitatory synaptic coupling strength and excitatory-inhibitory synaptic coupling strength) is required for the network to be able to exhibit stable persistent neural activity. The authors investigate the effect of extracellular potassium on perturbing this balance. They demonstrate that an increase in extracellular [K$^{+}$] narrows the region in the excitatory-inhibitory coupling strength parameter space, wherein the network can exhibit stable persistent neural activity. Furthermore there is an increase in the overall excitation in the network making the network more prone to exhibit seizure-like activity. These results are consistent with experimental findings of Rutecki et al. \cite{Rutecki:1985fu} wherein it was observed that raising extracellular [K$^{+}$], increased the the rate of spontaneous epileptiform discharges in {\it in vitro} hippocampal slices. Furthermore,  in support of these findings, Vincent at al. \cite{Vincent2011} have recently showed that by changing the concentration  of extracellular potassium in a rodent hippocampal slice preparation, the slice is prone to spontaneously transition to and from seizure-like states to regular activity. 

The authors use the above findings to then investigate the contrasting experimental findings related to
the glial contribution to epilepsy. It has been shown by Oberheim et al. \cite{Oberheim:2008bs} that the [Ca$^{2+}$] dependent
glutamate release from glial cells appear to synchronize the activity of adjacent neurons through simultaneous
non-synaptic slow inward neuronal current. The synchronized neuronal activity is then manifested in the
form of epileptic seizures. However, Fellin et al. \cite{Fellin:2006fv} report that the glutamate release by glia is not necessary for the generation of epileptic activity in hippocampal slices. Ullah et al. \cite{Ullah2009} use their neuronal network model to identify conditions under which the glutamate release by glial cells would cause their network  to exhibit seizure like activity. In the context of their network model, the glutamate release by glia is manifested in terms of transient increase in the excitatory-excitatory synaptic coupling within the network. They show that the effect of glia induced perturbations is dependent on the baseline level of excitatory-excitatory synaptic strength in the network, with higher baseline value resulting in higher likelihood for the network to generate seizure like activity in response to glia induced perturbations. Based on this findings the authors suggest that the experimental findings of Oberheim et al \cite{Oberheim:2008bs} and Felin et al \cite{Fellin:2006fv} can be explained in terms of different levels of baseline excitation under which the effect of glia induced perturbations were experimentally studied.

The second case example is the work by Santhakumar et al \cite{Santhakumar2005}, wherein the authors investigate the impact of mossy cell loss and mossy fiber sprouting on post traumatic excitability in the dentate gyrus (DG) subfield of the hippocampus. The modeling framework in this work is based on a biophysically detailed representation of the DG network using multi-compartmental neuron models to capture the morphology of the major DG cell populations: dentate granule cells, mossy and basket cells and the hilar perforant-path associated cells. Two types of DG-network architectures were investigated: a topographic network constrained by the axonal arborization pattern of each cell and a non-topographic network in which this limitation was relaxed. Detailed simulation studies using the above network model for DG showed that increasing the degree of mossy fiber sprouting in a non-topographic network resulted in propagation of activity from the directly activated granule cells to the other cells in the network. The degree of this propagation was proportional with the degree of the mossy cells sprouting and failed to sustain for longer periods of time. In contrast, in a topographic network, increased mossy fiber sprouting resulted in the faster propagation of activity to the entire network which eventually translates into self-sustained seizure-like network activity. This regime of network behavior was robust against a number of network parameters including synaptic conduction delay and synaptic strength. Further investigations into the relationship between the number of postsynaptic targets for the sprouted mossy fibers and the sustained quality of network activation revealed that there is an optimal pool of postsynaptic granule cells that correlates with self-sustained network activity. The study concludes by suggesting that the restricted topography of sprouted mossy fibers (which have been reported in experimental studies \cite{Buckmaster1999}) may play a central role in determining the spread of network activity in the DG.

It should be noted that other topographical features of network connectivity in DG may significantly impact the DG-network dynamics. For instance, Morgan and Soltez showed in a recent study \cite{Morgan2008} that the presence of highly interconnected neurons (hubs) in the DG network increases over all excitability within the DG-network making it prone to exhibit sustained seizure-like network activity.

Multi-compartment conductance based neuronal modeling have been used in earlier works to investigate mechanisms involved in epilepsy. Most notable are the studies performed by Traub and collaborators using a 19 compartment model for hippocampal CA1 pyramidal cells. Following a series of combined experimental/theoretical investigations \cite{Traub2002,Traub2005,Traub2005a,Traub2011} the authors conclude that  gap junctions between the axons of pyramidal neurons may play a critical role in the epileptic seizures with focal hippocampal origin. In other studies \cite{Traub1994a}, the authors have found that enhanced NMDA conductance can explain epileptiform activity observed in conditions of low extracellular [Mg$^{2+}$] in hippocampal slices.

The third case example is the study by Suffczynski et al. \cite{Suffczynski2004} wherein the authors investigate whether normal brain sleep-spindle activity and the pathological spike-wave discharges observed in epileptic brain tissue share a common underlying network mechanism, specifically each produced by interaction between the cortex and the thalamus. Since the focus of their investigation is interaction between multiple brain regions, they adopt the macro-scale modeling framework in their study. Specifically their network model was designed to capture the mean field activity of populations of neurons in the cortex and the thalamus. Their network was subjected to 3 different inputs: cortical input (received by the cortical pyramidal (PY) cells), sensory input (received by the thalamo-cortical (TC) cells) and an the third input received by the reticular thalamic cells. For a reference set of parameters, they show that the model exhibits bistable dynamical characteristics with a fixed point attractor coexisting with a limit cycle attractor. They suggest that the normal sleep-spindle oscillations corresponds to noise induced fluctuations in the network dynamics around the fixed point attractor of the network dynamics, whereas pathological spike-wave discharges correspond to the network dynamics evolving to a limit cycle attractor in the network.  Systematic bifurcation analysis of the network dynamics using the strength of cortical input $P_{ex}$ as the system parameter showed that the network dynamics transitions from a fixed point attractor to limit cycle attractor when the strength of cortical input exceeds a critical threshold $P_{ex}^{bif}$. Subsequently as P$_{ex}$ is gradually decreased, the network dynamics transitions to fixed point attractor for values of P$_{ex}<P_{ex}^{*}<P_{ex}^{bif}$. This analysis suggests that for values of cortical input $P_{ex}^{*}<P_{ex}<P_{ex}^{bif}$, the network dynamics is bistable and depending on the initial conditions the network dynamics can evolve to either normal fixed point attractor or the pathological limit cycle attractor. In this situation, noise can induce random transition of the network dynamics between the two states. The authors found that in this case, the  distribution of the duration of network dynamics in each of the two states is exponential, which echoes the experimental results on the duration of normal and paroxysmal epochs found in \cite{Bouwman2003}.

The authors also investigate the possibility of controlling the pathological spike-wave activity. They first show that a well timed external stimulus pulse of 40 Hz applied as cortical input for 10 ms caused the network to exhibit pathological spike-wave oscillations. They then show that a counter stimulus of same intensity applied at a specific phase of  the spike-wave oscillations can destroy this activity and restore the network activity to normal sleep-spindle state. These findings have potential applications in open-loop control strategies which will be discussed in Section 4.

It is evident from the case examples presented above that for many computational studies addressing possible epileptic mechanisms, enhanced excitability is a key factor which can drive a network towards an abnormal epileptiform seizure activity. However, recently there have been indications that higher degree of excitation relative to inhibition may not be a necessary condition for the manifestation of the epileptiform activity \cite{Cossart2005}. We would like to point the readers to an interesting study by Drongelen et al. \cite{Drongelen2005} wherein the authors systematically investigate the implications for this idea in a computational model of epileptic cortical network. Guided by an earlier study by Vreeswijk et al. \cite{Vreeswijk1994}, wherein the authors show that dependent on the time scale of synaptic interactions, inhibition can play a significant role in enhanced synchrony in  neuronal networks, the authors show that their cortical network model can exhibit emergent epileptiform activity in conditions of weak excitatory synapses. The authors have further tested the prediction of their modeling results in 
 mouse neocortical slices, by showing that a pharmacological reduction of excitatory synaptic transmission elicited sudden onset of repetitive epileptiform bursting behavior in the network.

\subsection{Non-deterministic Models}
The discussion thus far has focused on deterministic dynamical systems models for epilepsy, which is the primary workhorse in computational epilepsy modeling \cite{Lytton2008}. There are yet another category of models for epilepsy that fall under the umbrella of non-deterministic models. These models are based on the assumption that the observed brain dynamics is the result of a non-deterministic, high-dimensional dynamical system. Two broad category of non-deterministic models have found applications in epilepsy. Statistical models, wherein the focus has been on identifying statistical patterns in brain signals to guide the development of seizure prediction and detection algorithms and stochastic models, wherein the focus has been on the development of probabilistic models that can capture the apparent random transition of brain into and out of an epileptic seizure state.

Statistical models attempt to extract information embedded in brain recordings in order to develop predictors for seizure occurrence. Many statistical models employ nonlinear time series measures such as correlation dimension \cite{Elger:1998ly,Osorio:2001ys} and Lyapunov exponent \cite{Lehnertz:1999zr,Iasemidis:2005ve}  to identify patterns in the recorded brain signals indicative of an impending epileptic seizure.  Further more, tools from statistical learning theory such as support vector machines \cite{PredSeizVM1,PredSeizVM2}, Bayesian statistics \cite{PredSeizBayesian1,PredSeizBayesian2,PredSeizBayesian3} and artificial neural networks \cite{PredSeizANNGen1,PredSeizANNGen2} have been employed to enhance the efficacy of seizure prediction algorithms. A detailed exposition on statistical approach to computational modeling of epilepsy, with particular emphasis on applications to seizure prediction can be found in \cite{Iasemidis2003,Litt2002}.

Clinical investigations have found that seizure occurrence in epileptic patients follow complex patterns, ranging from periodic nature of seizure recurrence to situations when seizures are clustered around a particular time of the day to cases wherein there is no apparent identifiable timing pattern to seizure recurrence \cite{Milton:1987mi}. Stochastic models called Markov models and hidden Markov models have been proposed to explain and predict the existence of these complex seizure patterns \cite{Sunderam2001, Wong2007, Chiu2011}. The basic idea is that there are multiple attractor states within the brain. The transition between these attractor states is assumed to follow Markov property, namely the the transition of the brain to any future attractor state is only dependent on the present attractor state of the brain.  Based on this assumption, a probabilistic rule is identified that transitions the brain across various attractor states.  This modeling approach has found applications for both seizure prediction \cite{Sunderam2001} and also to assess the performance of statistical model based seizure prediction algorithms \cite{Wong2007}.

\section{Applications of computational models for epilepsy therapy}
Following from the discussion above on various computational modeling approaches in epilepsy, one might be left with the impression that deterministic-dynamical models are more appropriate for probing at the cellular and network mechanisms implicated in epilepsy whereas non-deterministic stochastic or statistical models are geared towards practical applications with focus on predicting the timing of seizure recurrence. While this is true to a certain extent, efforts are currently underway within the community of computational epilepsy researchers to bridge these two somewhat distinct modeling approaches. The drive in this direction is in part due to the recent emergence of control engineering approaches for the treatment of epilepsy \cite{Good:2009kx,Stigen:2011vn}. Such applications require both an excellent ability to predict an impending  seizure as well as a precise understanding of the mechanisms involved in seizure generation in order to develop control protocols to achieve long term seizure free status.

In the following sections, we will discuss some of the recent advances in modeling efforts that aim to interface the two modeling approaches to develop novel treatment protocols for epilepsy with specific emphasis on the utility for brain stimulation techniques for seizure control.

\subsection{Electrical Stimulation}
There is a general consensus  within the epilepsy community that, despite pharmacological and surgical advances in the treatment of epilepsy, seizures cannot be controlled in many patients and there is a need for new therapeutic approaches \cite{Jacobs:2001fk}. To this end, a growing body of clinical research indicates that controlling seizures may be possible through direct (deep brain) and indirect (vagus nerve) electrical stimulation  \cite{Theodore:2007uq}.
The initial success has propelled further research aimed towards improving the efficacy of this form of treatment for epilepsy. In particular, the focus has been on questions such as: what are the best brain structures to stimulate? and what is the most effective stimulation protocol? For example, it is not yet clear why high frequency and low frequency electrical stimulation have contrasting effects in different seizure models and structures \cite{Yamamoto:2002qf,Hamani:2004bh}. Computational models offer the potential to address some of these questions via systematic exploration through simulation experiments of varying electrical stimulation protocols across various brain regions.

The work of Tass and collaborators represents a step in this direction. Using an abstract network model of coupled oscillators, the authors propose a novel electrical stimulation protocol to suppress seizure like synchronous activity within the network. They hypothesize that effective electrical stimulation protocols facilitate the network to ``unlearn" the abnormal synchronized regime associated with epileptic seizures by means of synaptic plasticity mechanisms \cite{Abbott:2000pi}. Following from this hypothesis they show that high frequency pulse trains of electrical stimuli applied in coordinated fashion at different points within the network is able to suppress synchrony within the oscillator network. Furthermore, robust suppression of neural synchrony was maintained by using a closed-loop feedback signal that controlled the timing of application of the stimulus train and the width of applied stimulus pulses. The authors further investigated the effects of delayed feedback stimulation  to maintain desynchrony within the network \cite{Tass2005}. An important consideration of these research findings from the point of view of applications in seizure control is that the stimulation parameters are dynamically modulated depending on the state of brain activity without the need for time-consuming calibration of the stimulation parameters. The findings from above theoretical investigations were recently validated in an experimental study \cite{Tass2009} wherein sustained neural desynchrony was maintained in an epileptic hippocampal brain tissue via multisite coordinated feedback electrical stimulation.  Together, these studies offer a glimpse of the potential for computational models to aid in the design and implementation of novel electrical stimulation protocols that are highly effective in suppressing or even eliminating epileptic seizures.

Recently, clinical trails have been conducted to determine the efficacy of closed-loop electrical stimulation approach for seizure control \cite{Sohal2011,Morrell2011}. In this approach an online time series based statistical measure is employed to identify signatures of seizure or pre-seizure epileptic activity in the patient's EEG signal.
Following a successful detection of abnormal brain activity, a train of electrical stimulation pulse is delivered to the brain in order to suppress the occurrence of an impending epileptic seizure. Evidently, the success of this approach not only depends the specific stimulation protocol applied but more importantly on the detection and classification accuracy of the time series measure to detect abnormal brain activity. From this perspective advances in models for seizure prediction are expected to play an important role in enhancing the efficacy of closed-loop stimulation protocols.

\subsection{New direction: Light stimulation for seizure control}
Optogenetics is an emerging technology that leverages techniques from molecular biology, virology and genetic engineering to selectively express light sensitive ion channels in the membranes of either excitatory or inhibitory neurons \cite{Kuehn_2010}. It uses light to specifically excite [using algae protein channelrhodopsin-2 (ChR2)] or suppress [using light driven chloride pump halorhodopsin from archaea Natronomonas pharaonis (NpHR)] impulse activity in neurons with high degree of spatial and temporal resolution \cite{Boyden:2005dq} . As a result, this technique holds tremendous potential for fine external control of activity states in neuronal networks \cite{Zhang2007}. Furthermore, the channel kinetics of the light activated ion channels are better understood \cite{Hegemann:2005cr, Varo:1995nx} and amenable to mathematical modeling \cite{Nikolic:2009oq}. These models can be integrated into the Hodgkin Huxley conductance based neuron models \cite{Talathi:2011kl} and thus provide a natural framework suited to a computational modeling studies of the effect of light stimulation on brain networks and thereby offer the potential to design control strategies aimed at controlling epileptic seizures via light based stimulation protocols.

\begin{figure}[htbp]
  \begin{center}
    \includegraphics[scale=0.5]{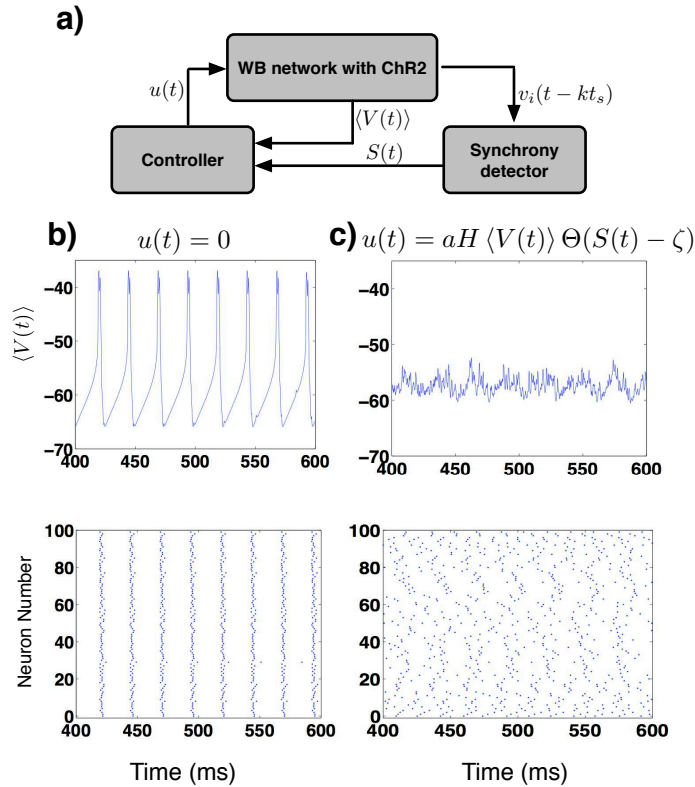}
  \caption{a) Schematic of closed-loop control architecture b) The mean field network activity (top) and the raster plot of neuronal spiking (bottom) in absence of control c) The response of the network when closed loop controller is active. Light intensity, I=0.01 mW/mm2}
  \label{Fig8}
  \end{center}
\end{figure}

Our research has recently focused on the question of how we can leverage the temporal and spatial precision of light stimulation to achieve robust suppression of pathological neural synchrony in brain networks? Here, we will present  results demonstrating the success of light stimulation based feedback controllers in achieving neural de-synchrony in a Wang-Buzsaki (WB) neuronal network of 100 all-to-all coupled identically firing interneurons \cite{Wang:1996tg}. WB network was chosen as an ideal template for our study because it is a classic model for neural synchrony in a biophysically realistic neuronal network. We modify the WB-network such that the membrane dynamics of each neuron in the network now involves an additional ion channel, the light sensitive protein ChR2. We further assume that the network is arranged in a ring geometry in a 2-dimensional Euclidean space. Neural synchrony was quantified using a synchrony metric $S(t)$ that has been well-characterized in literature \cite{HANSEL:1992hc}.  In absence of light stimulation, the WB network exhibits robust neural synchrony \cite{Wang:1996tg}. The stable synchronous firing state of the network is shown in Figure \ref{Fig8}b. We then tested the following linear proportional feed-back controller using light stimulation with low intensity such that light itself did not evoke an action potential in a given neuron in the network: $u(t)=aH\left<V(t)\right>\Theta\left(S(t)-\zeta\right)$, where $\left<V(t)\right>=(100 T)^{-1}\int_{t-T}^{t}du\sum_{i=1}^{100}v_{i}(u)$, $\Theta$ is the heaviside step function and $a$, $T$ and $\zeta$ are the control parameters. The light intensity was set at a nominal value of $H=0.01$ mW/mm$^{2}$. The schematic of the closed-loop control architecture is shown in Figure \ref{Fig8}a and in Figure \ref{Fig8}c, we present an example of successful closed loop control (suppression) of neural synchrony using the above linear proportional feed-back controller. This proof-of-principle example illustrates the potential of a weak light intensity stimulation based feedback controller for suppressing neural synchrony. We envision that in the future multi-disciplinary collaborative efforts between scientists with expertise in clinical epilepsy, molecular biology, computational modeling and control engineering will pave the path for the development of control systems and algorithms specifically designed to suppress pathological neural synchrony such as epileptic seizures originating in focal brain areas using spatially and temporally precise light stimulation protocols.

\section{Discussion}
The basic premise of this article is that epilepsy is a dynamical disease. Motivated by this idea, in this review article we have attempted to provide an uninitiated reader a brief introduction to the dynamical systems framework and then present examples from recent literature wherein concepts from dynamical systems are used to formulate computational models both at the micro-scale and the macro-scale to explore various mechanisms in epilepsy. We also present a brief discussion on computational models in epilepsy that fall under the general category of non-deterministic models. These models are primarily focused on addressing practical questions related to the predictive nature of seizure occurrence and methods aimed at assessing the performance of computational models for seizure prediction. We have also explored some applications, that have great potential for providing novel avenues for epilepsy treatment. In particular, we discuss recent progress in electrical stimulation based treatment protocols and also present preliminary results on the utility for light stimulation based protocols for controlling pathological brain activity.

As seen from the varying examples presented in this review, there are currently many well known mechanisms which can contribute to epileptic seizures and many computational models that explain how these mechanisms contribute to enhanced brain excitability leading to epileptic seizures. Despite these advances,  developing effective treatment protocols for patients suffering from refractory epilepsy has proven to be difficult. The primary reason for this undesirable situation is the vast complexity of the human brain and the epileptic syndrome. Most models, thought not simple,  are only able to capture a small portion of this complexity. However, we believe that the future is promising. Several investigations over the last 20 years focused on understanding the biology of the human brain has led to a wealth of data such as the human genome, the Allen atlas, human proteomics data base and the brain connectome. It is expected that the focus of future modeling efforts in epilepsy will make use of this great wealth of data to improve upon the existing models, address gaps in our knowledge, generate new predictions and possible provide avenue for new and effective treatment strategies.

Recent advancements in computer technology, including the availability of super computers (such as the Blue Gene), and new computational methods (such as distributed parallel computing architecture) are already allowing scientists to reverse engineer the brain to the molecular scale (the Blue Brain project) with the idea of obtaining better understanding of normal and abnormal brain function. Furthermore, advances in experimental technologies such as optogenetics are making it possible for us to achieve precise control of brain function at neuronal level thereby providing novel means to target and control brain networks susceptible to epileptic seizures. In this context, there are reasons to believe that more efficient treatment strategies for epilepsy are on the horizon.

In summary, in this review paper we present a brief survey for the utility for computational models in epilepsy. Our goal is to increase the awareness of computational modeling as one of the many tools at the disposal of epilepsy researchers, which can enable them to tackle the challenging problem of epilepsy and seizure control. Furthermore we hope that this review will enable collaborations between experimental and computational researchers that will ultimately result in more efficient treatment protocols for epilepsy.
\newline\newline
{\bf Acknowledgement}\\
SST would like to acknowledge kind support from the Wilder Center of Excellence for Epilepsy Research and the ChildrenÕs Miracle Network. We would also like to thank the anonymous reviewers for their critical feedback that helped us to significantly improve the manuscript.



\end{document}